\documentclass[a4paper,fleqn,usenatbib]{mnras}

\usepackage{newtxtext,newtxmath}

\usepackage[T1]{fontenc}
\usepackage{ae,aecompl}

\usepackage[dvipdfmx]{graphicx}	% Including figure files

\usepackage{ulem}
\usepackage{color}

\newcommand{\fdir}{./}
\newcommand{\msun}{M_\odot}
\newcommand{\rsun}{R_\odot}

\newcommand{\lsun}{L_\odot}
\newcommand{\mzams}{M_{\rm zams}}
\newcommand{\mc}{M_{\rm c}}
\newcommand{\mt}{M_{\rm t}}
\newcommand{\qcrit}{q_{\rm He,crit}}
\newcommand{\rpost}{R'}
\newcommand{\rcore}{R'_{\rm c}}
\newcommand{\yrgpc}{{\rm yr}^{-1}~{\rm Gpc}^{-3}}
\newcommand{\mi}{m_{\rm 1,i}}
\newcommand{\qi}{q_{\rm i}}
\newcommand{\ai}{a_{\rm i}}
\newcommand{\ei}{e_{\rm i}}
\newcommand{\mbp}{m_{\rm b,p}}
\newcommand{\mbs}{m_{\rm b,s}}
\newcommand{\pmass}{85^{+21}_{-14}M_\odot}
\newcommand{\smass}{66^{+17}_{-18}M_\odot}
\newcommand{\tmass}{150^{+29}_{-17}M_\odot}
\newcommand{\rshft}{0.82^{+0.28}_{-0.34}}
\newcommand{\rrate}{0.13^{+0.30}_{-0.11}}

\title[Pop.~III Binary Black Holes]{Population III Binary Black Holes:
  Effects of Convective Overshooting on Formation of GW190521}

\author[Ataru Tanikawa]{
Ataru Tanikawa,$^{1,}$\thanks{E-mail: tanikawa@ea.c.u-tokyo.ac.jp}
Tomoya Kinugawa,$^{2}$ Takashi Yoshida,$^{3}$ Kotaro Hijikawa,$^{3}$
Hideyuki Umeda$^{3}$
\\
$^{1}$Department of Earth Science and Astronomy, College of
  Arts and Sciences, The University of Tokyo, 3-8-1 Komaba, Meguro-ku,
  Tokyo 153-8902, Japan\\
$^{2}$Institute for Cosmic Ray Research, The University of Tokyo,
  Kashiwa, Chiba, Japan\\
$^{3}$Department of Astronomy, Graduate School of Science, The
  University of Tokyo, Bunkyo-ku, Tokyo, Japan}

\date{Accepted XXX. Received YYY; in original form ZZZ}

\pubyear{2017}

\begin{document}
\label{firstpage}
\pagerange{\pageref{firstpage}--\pageref{lastpage}}
\maketitle

\begin{abstract}

GW190521 is a merger of two black holes (BHs), wherein at least one BH
lies within the pair-instability (PI) mass gap, and it is difficult to
form because of the effects of PI supernovae (PISNe) and pulsational
PI (PPI). In this study, we examined the formation of GW190521-like
BH-BHs under Population (Pop) III environments by binary population
synthesis calculations. We reveal that convective overshooting in
stellar evolution strongly affects the formation of GW190521-like
BH-BHs. A model with a small overshoot parameter (similar to GENEC)
can form GW190521-like BH-BHs. The derived merger rate is $4 \times
10^{-2}$~$\yrgpc$ at a redshift of $\sim 0.82$, which is comparable to
the merger rate of GW190521-like BH-BHs inferred by gravitational wave
(GW) observations. In this model, a $\sim 90~\msun$ star collapses to
form a $\sim 90~\msun$ BH by avoiding PPI and PISN even if it is a
member of a binary star. This is because it expands up to
$10^2~\rsun$, and lose only little mass through binary
evolution. However, a model with a large overshoot parameter (similar
to Stern) cannot form GW190521-like BH-BHs at all. Thus, we cannot
conclude that a Pop~III binary system is the origin of GW190521
because determination of the overshoot parameter involves highly
uncertain. If a Pop~III binary system is the origin of GW190521, the
merger rate of BH-BHs including a $100-135~\msun$ BH is substantially
smaller than that of GW190521-like BH-BHs. This will be assessed by GW
observations in the near future.

\end{abstract}


\section{Introduction}
\label{sec:Introduction}

The first binary black hole (BH-BH) merger was observed in the first
direct detection of gravitational waves (GWs)
\citep{2016PhRvL.116f1102A}. Since then, many BH-BHs have been
discovered. By the end of the first and second observing runs (O1 and
O2, respectively) 10 BH-BHs were discovered
\citep{2019PhRvX...9c1040A}. This number increased by 4 times in the
first-half of the third observing run (O3a)
\citep{2020arXiv201014527A, 2020arXiv201014533T}. The O3a has not only
discovered more BH-BHs, but has also discovered BH-BHs with different
properties from those discovered in the O1/O2, such as GW190412 with
asymmetric masses \citep{2020PhRvD.102d3015A} and GW190814 with a
compact object in the lower mass gap (MG) \citep{2020ApJ...896L..44A}.

GW190521 was also discovered in the O3a, which is a BH-BH with a total
mass of $150~\msun$ \citep{2020PhRvL.125j1102A}. The presence of
GW190521 is surprising because at least one of the two BHs can be in
the pair-instability (PI) MG $65-130~\msun$, which was not discovered
in the O1/O2 \citep{2019ApJ...882L..24A}. This type of BHs are
difficult to form because of PI supernovae (SNe)
\citep{1967PhRvL..18..379B, 1968Ap&SS...2...96F, 1984ApJ...280..825B,
  1986A&A...167..274E, 2001ApJ...550..372F,
  2002ApJ...567..532H,2002ApJ...565..385U}, and pulsational PI (PPI)
\citep{2002ApJ...567..532H, 2016MNRAS.457..351Y, 2019ApJ...887...72L}.
Many formation scenarios of GW190521 have been summarized by
\cite{2020ApJ...900L..13A}: hierarchical BH mergers in globular
clusters \citep{2019PhRvD.100d3027R}, stellar collisions in open
clusters \citep{2020MNRAS.497.1043D}, and combination of BH mergers
and gas accretion in disks of active galactic nuclei
\citep{2019PhRvL.123r1101Y, 2020ApJ...898...25T,
  2020arXiv200715022M}. Gas accretion may bridge the PI~MG
\citep{2019A&A...632L...8R}.  \cite{2020ApJ...904L..26F} have proposed
that GW190521 consists of BHs below and above the PI~MG. If so,
GW190521 can be formed through Population (Pop) I/II and Pop~III
binary evolutions
\citep[][respectively]{2019ApJ...883L..27M,2021ApJ...910...30T}. An
electromagnetic counterpart of GW190521 was actively discussed
\citep{2020PhRvL.124y1102G, 2021arXiv210316069P}.

\cite{2021MNRAS.502L..40F} (F21) have claimed that Pop~III binary
systems can form GW190521. This is because Pop~III stars lose little
mass through stellar winds, keep their stellar radii small so as not
to interact with their companion stars, and have small carbon-oxygen
(CO) core mass because of H-He shell interactions. This is contrasting
to the results obtained by \cite{2021ApJ...910...30T} (T21) in which
Pop~III binary systems hardly form GW190521-like BH-BHs. T21 have not
found the formation of GW190521-like BH-BHs. This discrepancy results
from the choice of a convective overshoot parameter for a stellar
evolution model. F21 and T21 have adopted small and large overshoot
parameters, respectively. If a convective overshoot parameter is
small, a Pop~III star keeps a small radius until collapse to a BH,
loses little mass through binary interaction and can leave BHs in the
PI~MG. It is difficult to identify the correct parameters. The
overshoot parameter in F21's model is the same as that in GENEC to
explain the main-sequence width of AB stars observed in open clusters
in the Milky Way Galaxy \citep{2012A&A...537A.146E}, while the
overshoot parameter in T21's model is calibrated based on Stern to
explain the early B-type stars observed in the Large Magellanic Cloud
\citep{2011A&A...530A.115B}.

\cite{2021MNRAS.501L..49K} (K21) have suggested that Pop~III binary
systems can form a large number of GW190521-like BH-BHs, and that the
merger rate is comparable to the rate of GW190521, despite that they
have adopted a stellar evolution model constructed by
\cite{2001A&A...371..152M} with similar behaviors to a model with a
large convective overshoot parameter. This is because of their
modeling of post main-sequence (MS) stellar radii. However, we show
that their modeling is not applicable to Pop~III stars that form BHs
in the PI~MG.

In this study, we showed that the choice of convective overshooting
parameters strongly affects the formation of GW190521-like BH-BHs from
Pop~III binary systems. We perform binary population synthesis (BPS)
calculations for three stellar models with different convective
overshoot parameters. Two of the three models have small and large
convective overshoot parameters, and are called the M and L models,
respectively \citep{Yoshida19}. The overshoot parameters of the M and
L models are calibrated to GENEC \citep{2012A&A...537A.146E} and Stern
\citep{2011A&A...530A.115B}, respectively. The M and L models are
named after the first letters of the Milky Way galaxy and the Large
Magellanic Cloud, whose stars are used for calibrations by GENEC and
Stern, respectively. The M model is similar to the F21's model. The L
model is the same as the T21's model. The third model is the L model
with K21's modeling of post-MS stellar radii, which is similar to the
K21's model.

\cite{2020MNRAS.495.2475L,2020ApJ...903L..40L} examined the formation
of GW190521-like BH-BHs from Pop~III stars through dynamical
interactions, and \cite{2020ApJ...903L..21S} considered Pop~III BH
growth via gas accretion. In contrast, we focus on formation of
GW190521-like BH-BHs from Pop~III binary systems through pure binary
evolution. \cite{2020ApJ...905L..15B} suggested that GW190521-like
BH-BHs can be formed from Pop~II binary systems through pure binary
evolution, if stars with He cores of $90~\msun$ can avoid PPI and
PISNe \citep{2018ApJ...863..153T, 2020ApJ...902L..36F,
  2021MNRAS.501.4514C}. This is true only if the
$^{12}$C$(\alpha,\gamma)^{16}$O reaction rate is lower than its
standard reaction rate by $3\sigma$. In this study, we assumed that
stars with He cores of $45-65~\msun$ experience PPI and leave
$45~\msun$ BHs, and that stars with He cores of $65-135~\msun$
experience PISNe, and leave no remnants. This assumption makes it
difficult to form GW190521-like BH-BHs through pure binary evolution;
however, it is based on the results of PPI and PISNe with the standard
$^{12}$C$(\alpha,\gamma)^{16}$O reaction rate. Many other formation
scenarios have been suggested immediately after the publication of
GW190521 \citep{2020arXiv200904360M, 2020arXiv200905461G,
  2020arXiv200910688P, 2020arXiv201006161A, 2020PhRvL.125z1105S,
  2020ApJ...902L..26F, 2020ApJ...903L...5R, 2020ApJ...904L..13R,
  2021ApJ...908...59R, 2021PhRvL.126h1101B, 2021PhRvL.126e1101D}.
After submission of this manuscript, we became aware of the
complementary work by \cite{2021MNRAS.504..146V}, which independently
finds that uncertainty in convective overshooting leads to significant
uncertainty in predictions for the maximum BH mass which can be formed
below the PI MG.

The remainder of this paper is structured as follows. In
section~\ref{sec:Method}, we present our {\tt BSE} code and initial
conditions. In section~\ref{sec:Results}, we detail the results of the
BPS calculations. In section~\ref{sec:Summary}, we summarize this
study.

\section{Method}
\label{sec:Method}

\begin{figure}
  \includegraphics[width=\columnwidth]{\fdir/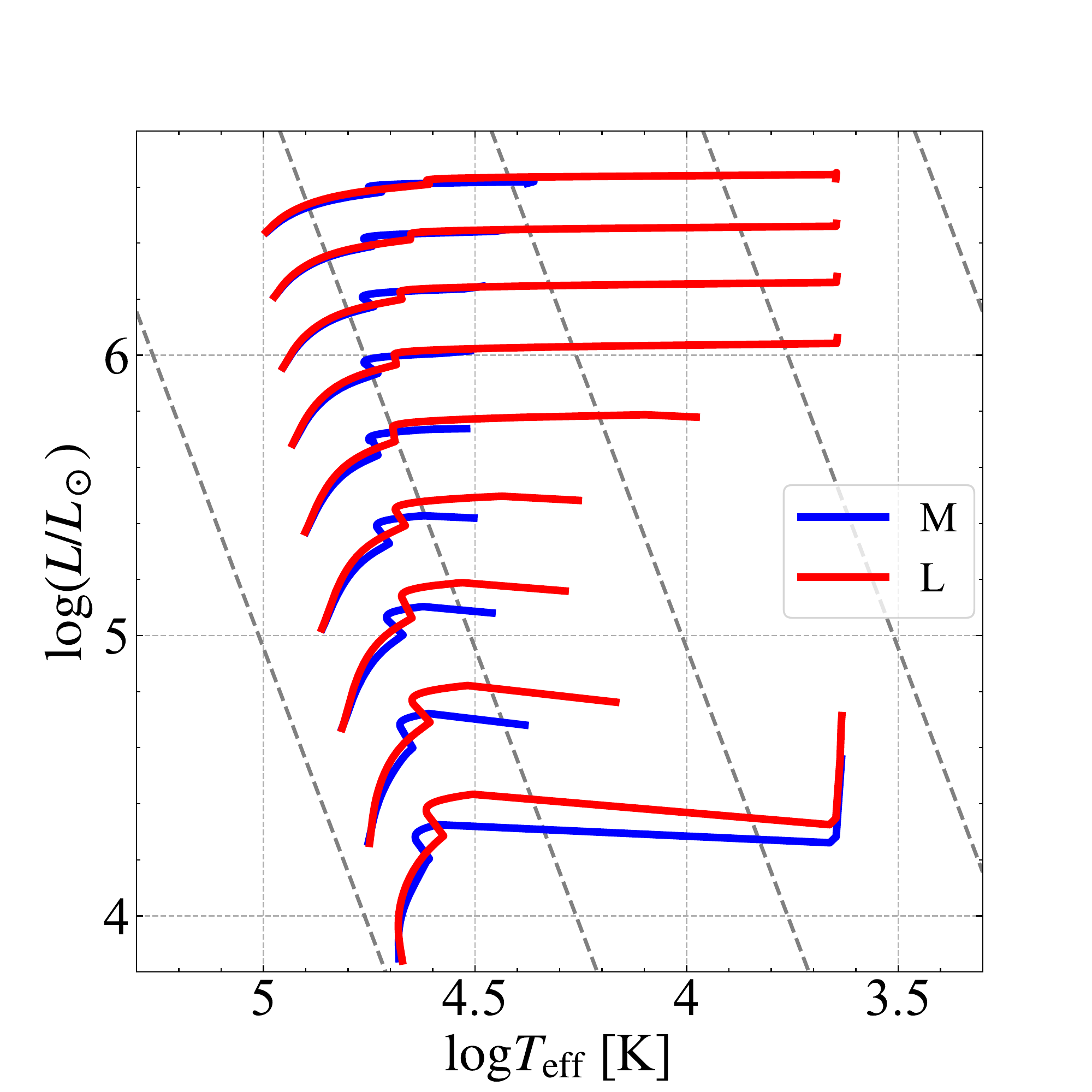}
  \caption{Hertzsprung--Russell (HR) diagram of stars with
    $\mzams=10-160~\msun$ at intervals of $2^{1/2}$ from bottom to top
    for the M (blue) and L (red) models. Gray dashed lines indicate
    stellar radii of $1-10^4 \rsun$ at intervals of $10$ from left to
    right.}
  \label{fig:hrd}
\end{figure}

We used a widely-used BPS code {\tt BSE} \citep{2000MNRAS.315..543H,
  2002MNRAS.329..897H} with extensions to massive and extremely
metal-poor stars \citep{2020MNRAS.495.4170T}. Herein, we briefly
introduce our {\tt BSE} code, described in detail in T21 and
\cite{2020MNRAS.495.4170T}. We did not account for stellar wind mass
loss because of Pop~III stars, although our {\tt BSE} code can include
stellar wind mass loss. Figure~\ref{fig:hrd} shows
Hertzsprung--Russell (HR) diagram of stars $\mzams=10-160~\msun$ at
intervals of $2^{1/2}$ from bottom to top for the M (blue) and L (red)
models. Stars in the M model maintain substantially smaller radii than
in the L model. This is because of difference in convective overshoot
parameters between these models. In our stellar evolution
calculations, convective overshoot is taken with a diffusive
treatment.  The diffusion coefficient exponentially decreases with the
distance from a convection boundary with a scale of $f_{\rm ov}
H_{P}/2$, where $H_{P}$ is the pressure scale height at the
boundary. The L model adopts larger convective overshoot parameter
($f_{\rm ov}=0.03$) than the M model ($f_{\rm ov}=0.01$). When
convective overshooting is more effective, post-MS stars have larger
He core masses. The larger He cores emit larger luminosities, and the
larger luminosities expand the stellar radii more
\citep[e.g.][]{1992PASP..104..717P}. In the L model, the star with
$\mzams = 80~\msun$ exceeds $10^3 \rsun$, which is consistent with the
results of \cite{2001A&A...371..152M} and
\cite{2012A&A...542A.113Y}. In the M model, the star with $\mzams =
80~\msun$ expands up to $\sim 40~\rsun$, which is smaller than the
radius of the corresponding radius in F21, $\sim
160~\rsun$. Nevertheless, this is rather helpful for the aim of
investigating the dependence of GW190521 formation on convective
overshooting.

We highlight that the M and L models are not the same as GENEC and
Stern, respectively. \cite{Yoshida19} did not compare their simulation
results with observed stars themselves. Instead, they showed that in
the solar metallicity the evolutions of $20~\msun$ stars in the M and
L models are comparable to those calculated by the GENEC and Stern
codes, respectively. These stars have remarkably higher metallicity
than Pop~III stars and have much smaller mass than stars forming BHs
in PI~MG. Moreover, the {\tt HOSHI} code treats convection differently
from the GENEC and Stern codes. Both the Stern and {\tt HOSHI} codes
use the Ledoux criterion for convection; however, they adopt different
semiconvection parameters. The GENEC code uses the Schwarzschild
criterion for convection. Moreover, the {\tt HOSHI} code uses an
exponential description of convective overshooting, while the GENEC
and Stern codes use step-function overshooting descriptions. Thus, the
M and L models do not have the same quantitative evolution of
$80~\msun$ stars with zero metallicity as the GENEC and Stern codes,
although they have qualitatively the same. We also mention our
treatment of stellar surface in the {\tt HOSHI} code. It is difficult
to solve the evolution of the convective envelope with high Eddington
factors in massive stars and some numerical treatments are adopted to
be solved \citep{2013ApJS..208....4P, 2017A&A...597A..71S}. The
different treatments would affect the radial evolution. In this study,
we did not adopt any special treatments in the {\tt HOSHI} code to
solve the surface evolution.  The surface evolution problem is less
serious in Pop III massive stars\footnote{We limit the opacity so that
  the local luminosity does not exceed the Eddington luminosity in the
  region where $M_r > 0.99 M_{\rm total}$ in case of $Z \ge 0.01$
  Z$_\odot$ star.}.

\begin{figure}
  \includegraphics[width=\columnwidth]{\fdir/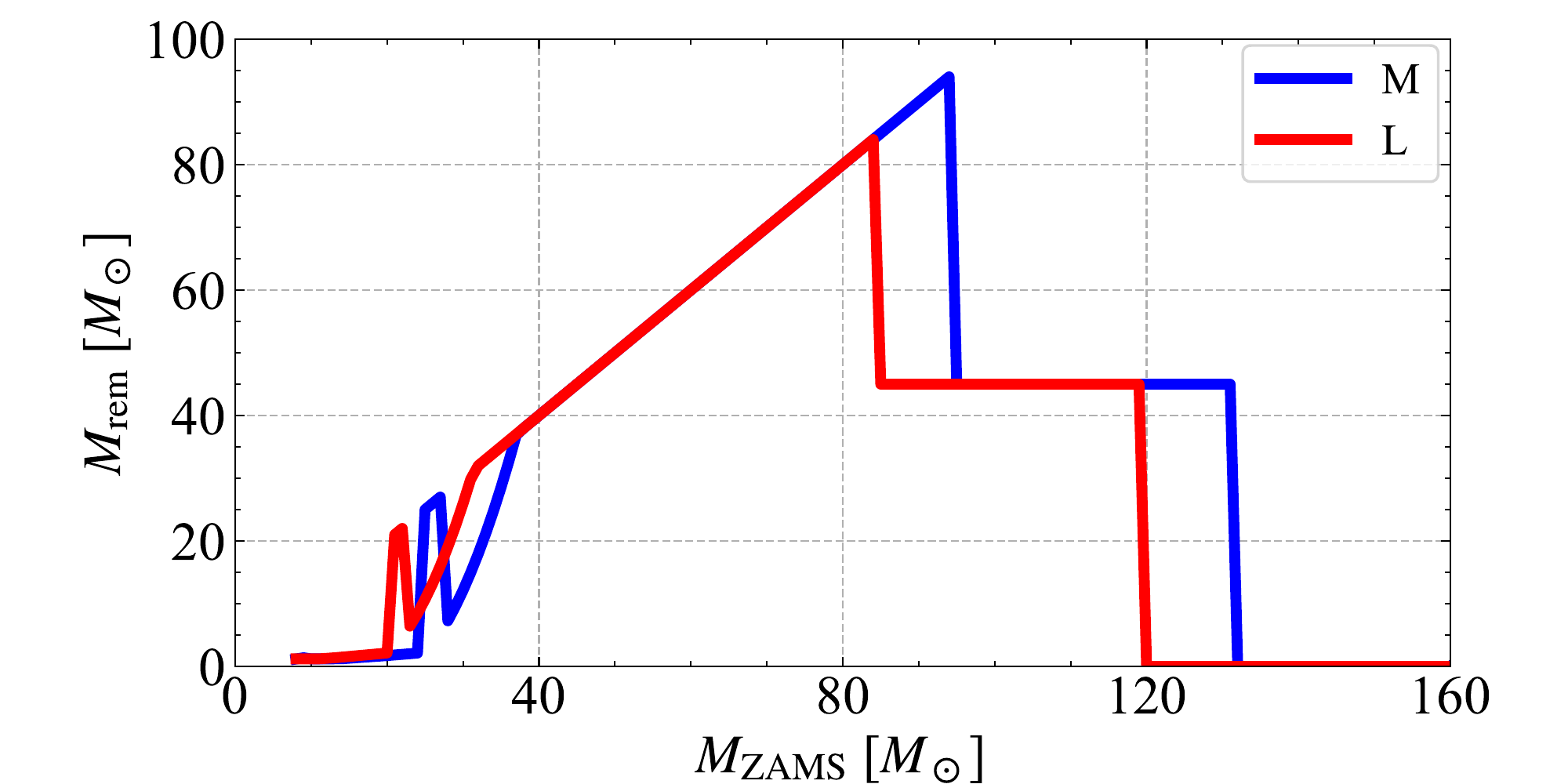}
  \caption{Relation between zero-age main-sequence (ZAMS) and remnant
    masses in the M and L models.}
  \label{fig:remnantMass}
\end{figure}

For a supernova model, we adopted the rapid model in
\cite{2012ApJ...749...91F} modified by PPI and PISNe. Stars with He
core masses of $45-65~\msun$ and $65-135~\msun$ experience PPI and
PISNe, respectively. If stars experience PPI and PISNe, they leave
$45~\msun$ BHs and no remnants, respectively, which is similar to a
strong PPI model constructed by \cite{2016A&A...594A..97B}\footnote{
  \cite{2020ApJ...905L..21U} have intensively investigated PPI and
  PISN of Pop~III stars.}. Figure~\ref{fig:remnantMass} shows the
relation between ZAMS and remnant masses in the M and L models. In
both the models, stars with $\mzams \sim 90 - 130~\msun$ leave
$45~\msun$ BHs because of PPI, and stars with $\mzams \gtrsim
130~\msun$ leave no remnants because of PISNe. Stars with $\mzams \sim
50 - 90~\msun$ can leave BHs with $\gtrsim 45~\msun$, because they
have small He core mass with $\lesssim 45~\msun$, and do not
experience PPI nor PISNe. \cite{2020ApJ...888...76M} and
\cite{2021MNRAS.501.4514C} have also predicted BHs with $\gtrsim
45~\msun$ can be formed through the same mechanism.  The L model has
lighter ZAMS stars undergoing PPI and PISNe than the M model. This is
because stars form larger He cores in the L model than in the M
model. Furthermore, we do not consider BH natal kicks.

Our {\tt BSE} code treats binary interactions similar to the original
{\tt BSE} code \citep{2002MNRAS.329..897H}. Here, we describe
different points between the original and our {\tt BSE} code. The
original {\tt BSE} code considers the stars with convective and
radiative envelopes as those in the phases of core helium burning and
shell helium burning, respectively. In contrast, our {\tt BSE} code
models stars with convective and radiative envelopes as stars with the
effective temperature of $< 10^{3.65}$~K and $> 10^{3.65}$~K,
respectively. This affects the tidal interactions and mass
transfer. In tidal interactions, the equilibrium tide with convective
damping is adopted for stars with convective envelopes, while the
dynamical tide with radiative damping is adopted for stars with
radiative envelopes. In mass transfer, stars with convective envelopes
experience unstable mass transfer (common envelope evolution) more
easily than those with radiative envelopes. Common envelope evolution
is approximated as the $\alpha$ formalism
\citep[e.g.][]{1984ApJ...277..355W}. We adopt $\alpha_{\rm CE}
\lambda_{\rm CE}=1$, where $\alpha_{\rm CE}$ is the common envelope
efficiency, and $\lambda_{\rm CE}$ is the structural binding energy
parameter. Furthermore, we consider orbital shrinkage because of GW
radiation.  We switch off magnetic braking because Pop~III binary
systems should have only tangled magnetic field
\citep{2020MNRAS.497..336S}.

We prepare $10^6$ binary systems for initial conditions. Their
distributions of initial primary masses ($\mi$), mass ratios $\qi$,
semi-major axes ($\ai$), and eccentricities ($\ei$) are the same as
those of the optia1e1q0.0 model in T21 except for the maximum primary
mass. The details are as follows. The distribution of the initial
primary mass ($\mi$) is proportional to $\mi^{-1}$ in the range of $10
- 150~\msun$ because the initial mass function of Pop~III stars are
predicted to be logarithmically flat in the mass range from a few
$10~\msun$ to a few $100~\msun$ by numerical simulations
\citep[e.g.][]{2014ApJ...792...32S, 2014ApJ...781...60H}. Pop~III
stars can have larger masses and our main objective is to form BHs in
the PI MG; thus, they are not considered here. The distribution of
initial mass ratio $\qi$ is uniform in the range from $10~\msun/\mi$
to $1$. The distribution of initial semi-major axis ($\ai$) is
logarithmically flat ($\propto \ai^{-1}$) in the range of $10 -
2000~\rsun$. We set the eccentricity distribution to the thermal
distribution, $\propto \ei$. We exclude binary systems if at least one
of two stars fills its Roche lobe at the initial time.

We applied the M and L models to these binary systems. The M model is
similar to the F21's model, and the L model is the same as the T21's
model. Additionally, we adopt the L model with K21's modeling of
post-MS radii (hereafter, L+K21 model), which is similar to the model
of K21. In the L+K21 model, stars evolve along with the L model unless
they experience stable mass transfer. Post-MS stars in the L+K21 model
respond to mass loss through mass transfer differently from post-MS
stars in the L model in the following. Generally, a post-MS star
slowly shrinks with mass loss of its hydrogen envelope, and steeply
shrinks to its He core size around when it completely loses its
hydrogen envelope. Thus, we modeled the response of a post-MS radius
to mass loss, such that a post-MS star suddenly shrinks down to its He
core size, when a fraction of its He core mass ($\mc$) to its total
mass ($\mt$) becomes larger than a critical value.  Hereafter, we call
the critical value $\qcrit$. In other words, a post-MS star shrinks to
its He core size when $\mc/\mt > \qcrit$. In the M and L models,
$\qcrit \sim 0.99$, which is the same as the {\tt SSE/BSE} codes (see
Appendix~\ref{sec:qcrit}). In contrast, in the L+K21 model, $\qcrit =
0.58$\footnote{Actually, K21 have not explicitly described that they
  have adopted $\qcrit=0.58$.}. This is based on the result of
\cite{2017MNRAS.468.5020I} (see their section 2.2.5).  This $\qcrit$
may be appropriate for Pop~III stars with $\mzams = 20 - 50~\msun$;
however, it may not be appropriate for Pop~III stars with $\mzams = 60
- 90~\msun$, which form BHs in the PI~MG
(Figure~\ref{fig:remnantMass}). We discuss this in
Section~\ref{sec:Results}.

We assumed a simple Pop~III star formation model as follows. All
Pop~III stars are formed in minihalos at the redshift of $\sim
10$. All the minihalos form one Pop~III binary systems. The number
density of minihalos is $10^{11}$~Gpc$^{-3}$, and then the number
density of Pop~III binary systems is $10^{11}$~Gpc$^{-3}$. This
formation model is consistent with \cite{2019MNRAS.487..486M} and
\cite{2020MNRAS.492.4386S} with respect to the total Pop~III mass in
the local universe. In contrast, the total Pop~III mass is much
smaller than estimated by \cite{2011A&A...533A..32D} and
\cite{2016MNRAS.461.2722I}, which are adopted by
\cite{2014MNRAS.442.2963K, 2020MNRAS.498.3946K, 2021PTEP.2021b1E01K},
and K21. Our Pop~III formation rate may be pessimistically small.

\begin{table}
  \centering
  \caption{Parameters of GW190521 we adopted.}
  \label{tab:GW190521}
  \renewcommand{\arraystretch}{1.3}
  \begin{tabular}{lll}
    \hline
    Parameter & & \\
    \hline
    Primary mass   & $\pmass$        & $90$\% credible intervals \\
    Secondary mass & $\smass$        & $90$\% credible intervals \\
    Total mass     & $\tmass$        & $90$\% credible intervals \\
    Redshift       & $\rshft$        & $90$\% credible intervals \\
    Rate &         $\rrate$~$\yrgpc$ & \\
    \hline
  \end{tabular}
\end{table}

Table~\ref{tab:GW190521} presents the parameters of GW190521 we
adopted. All parameters (except rate) have error bars of $90$\%
credible intervals. These parameters are detailed by
\cite{2020PhRvL.125j1102A} and \cite{2020ApJ...900L..13A}.

\section{Results}
\label{sec:Results}

\begin{figure}
  \includegraphics[width=\columnwidth]{\fdir/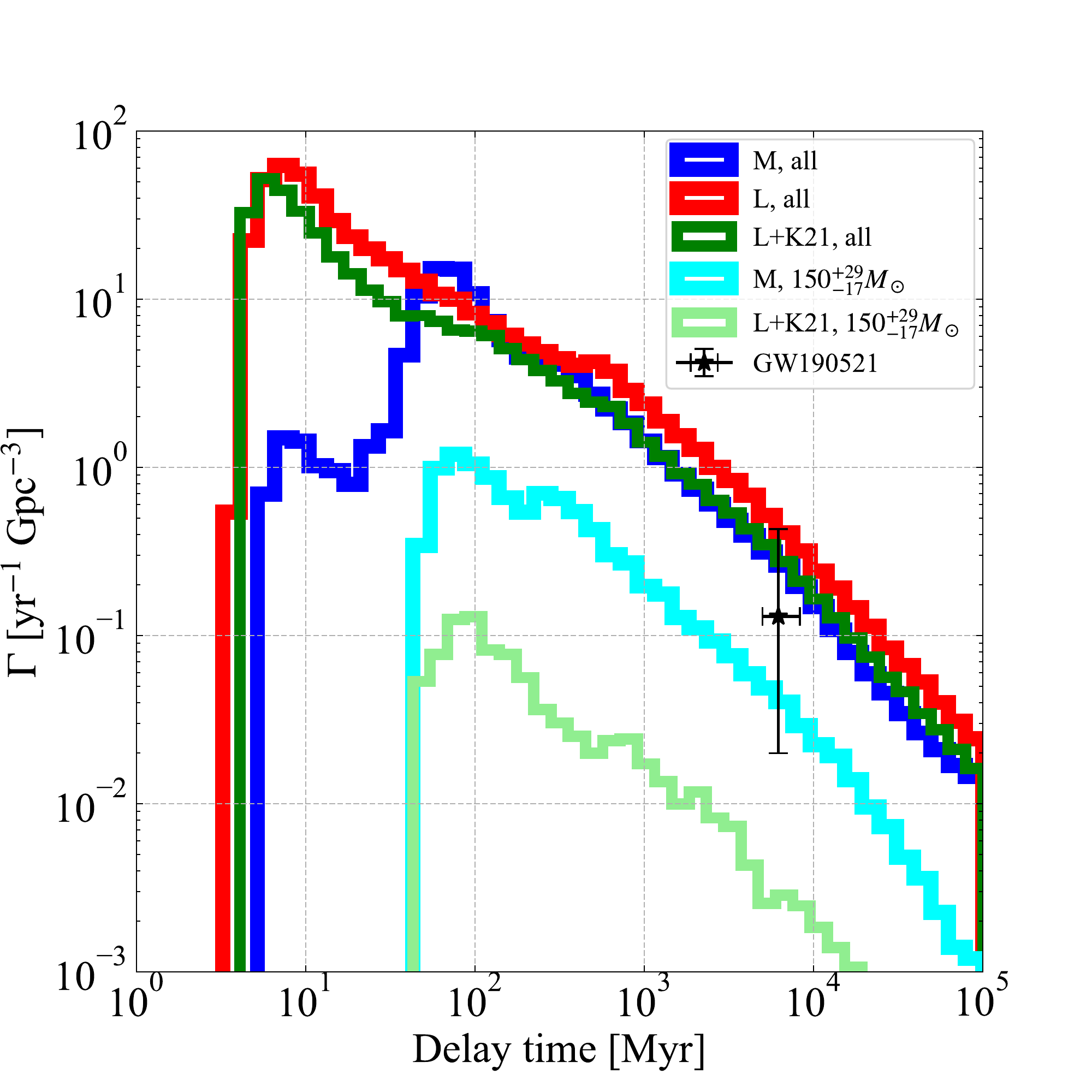}
  \caption{Merger rate density of Pop~III BH-BHs in the M (blue), L
    (red), and L+K21 (green) models as a function of delay time. The
    light-blue and light-green curves indicate BH-BH mergers in the M
    and L+K21 models, where their total masses are within the error
    bar of the total mass of GW190521. There is no such BH-BH mergers
    in the L model. The delay time and rate of GW190521 are indicated
    by the star mark with error bars.}
  \label{fig:delayTime}
\end{figure}

We define the merger rate density of BH-BHs as follows:
\begin{align}
  \Gamma = \frac{d(N_{\rm BH-BH}/N_{\rm bin})}{dt_{\rm d}} \left(
  \frac{\eta_{\rm bin}}{1} \right) \left( \frac{n_{\rm
      DM}}{10^{11}{\rm Gpc}^{-3}} \right) \; [{\rm yr}^{-1} \; {\rm
      Gpc}^{-3}],
\end{align}
where $N_{\rm bin}$ is the number of simulated binary systems, $N_{\rm
  BH-BH}$ is the number of merging BH-BHs, $t_{\rm d}$ is the delay
time, $\eta_{\rm bin}$ is the number of Pop~III binary systems formed
in each minihalo, and $n_{\rm DM}$ is the number density of the
minihalo.  Figure~\ref{fig:delayTime} shows the merger rate density of
BH-BHs in the M, L, and L+K21 models as a function of the delay
time. After $\sim 10^2$~Myr, the distributions of all BH-BH merger are
similar among the three models. In all the models, these BH-BHs are
formed through stable mass transfer. In contrast, the merger rates in
the L and L+K21 models are much larger than in the M model before
$\sim 10^2$~Myr. Stars expand up to $10^3 \rsun$ and evolve to form
red-supergiant stars in the L and L+K21 models; thus, binary stars can
experience common envelope evolution. The resulting BH-BHs can have
semi-major axes comparable to the size of He cores, and have short GW
radiation timescales ($\lesssim 10^2$~Myr). If binary stars do not
undergo common envelope evolution, their semi-major axes are at least
the size of their ZAMS radii, and then their GW radiation timescales
are long, $\gtrsim 10^2$~Myr. In all the models, the merger rates of
BH-BHs at $\sim 10$~Gyr are $\sim 10^{-1}$~yr$^{-1}$, which is
consistent with T21.

Figure~\ref{fig:delayTime} shows the merger rate density of BH-BHs,
whose total masses are $150^{+29}_{-17} \msun$, within the error bar
of the total mass of GW190521. There are many such BH-BHs in the M
model, while there is no such BH-BH in the L model. In the L model,
stars with $\mzams = 50-90~\msun$ expand to $\gtrsim 10^3~\rsun$, and
evolve to red-supergiant stars. Such stars interact with their
companion stars and experience large mass loss through stable mass
transfer or common envelope evolution. They lose their hydrogen
envelopes and leave naked He stars. If the naked He stars have
$45-65~\msun$ and $> 65~\msun$, they leave $45~\msun$ BHs and no
remnants, respectively. They can leave only BHs with $\lesssim
45~\msun$. Thus, they cannot form GW190521-like BH-BHs.  In the M
model, stars with $\mzams = 50-90~\msun$ expand up to $\sim
10^2~\rsun$, and keep blue-supergiant stars until they collapse to
BHs. Such stars lose little mass through stable mass transfer, and
tend not to experience common envelope evolution. They can maintain
massive hydrogen envelopes. They can leave BHs with $50-90~\msun$
because their He core masses are $\lesssim 45~\msun$. Eventually, they
can form GW190521-like BH-BHs. We discuss about the L+K21 model later.

We plot the delay time of GW190521 with error bars in
Figure~\ref{fig:delayTime}. We adopt $z=0.82^{+0.28}_{-0.34}$ for the
redshift of GW190521, and $0.13^{+0.30}_{-0.11}$~$\yrgpc$ for its
rate, as presented in Table~\ref{tab:GW190521}. The rate of BH-BHs
with total masses of $150^{+29}_{-17} \msun$ in the M model is
slightly smaller than the median value of the rate of GW190521-like
events, and within the error bars of $90$\% credible intervals.  Our
Pop~III formation model is pessimistically small; therefore our rate
can be regarded as the lower limit. Thus, a Pop~III binary can be the
origin of GW190521, if the M model can apply for Pop~III stars. In
contrast, if Pop~III stars evolve along with the L model, no Pop~III
binary can form GW190521-like events.

\begin{figure}
  \includegraphics[width=\columnwidth]{\fdir/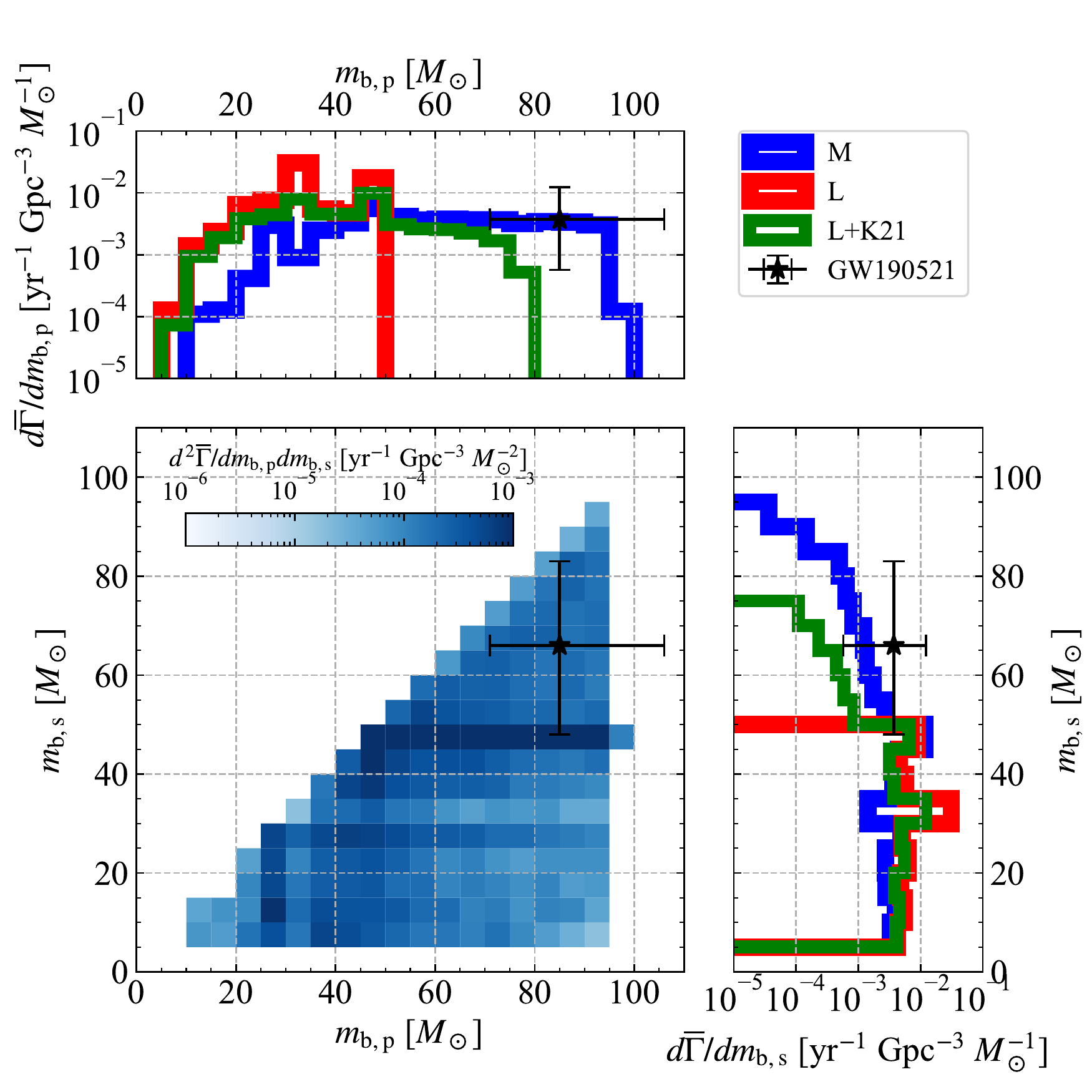}
  \caption{Mass distribution of the merging BH-BHs at the redshift of
    the GW190521 event in the M (blue), L (red), and L+K21 (green)
    models. The $\overline{\Gamma}$ is defined in the main text, and
    $\mbp$ and $\mbs$ are the primary and secondary BH masses
    respectively, where $\mbp \ge \mbs$. The 2D histogram indicates
    just the M model. The mass and rate of GW190521 are indicated by
    the star marks with error bars. In order to obtain the rate per
    mass of GW190521, we divide the rate of GW190521 by the widths of
    error bars of the BH masses, assuming that the rate is distributed
    uniformly in the error bars of the BH masses.}
  \label{fig:massDist}
\end{figure}

Figure~\ref{fig:massDist} shows the mass distributions of BH-BHs in
the M, L, and L+K21 models (see Appendix~\ref{sec:NumericalNoise} to
quantify their numerical noise). We define the averaged merger rate
density $\overline{\Gamma}$ over the error bar of the delay time of
GW190521 as follows:
\begin{align}
  \overline{\Gamma} = \frac{1}{t_{\rm d,f}-t_{\rm d,i}} \int_{t_{\rm
      d,i}}^{t_{\rm d,f}} \Gamma dt_{\rm d},
\end{align}
where we adopt $t_{\rm d,i}=5.5$~Gyr and $t_{\rm d,f}=8.8$~Gyr. To
investigate the BH-BH mass distribution, we differentiate the averaged
merger rate by the primary BH mass $\mbp$ or secondary BH mass $\mbs$,
where $\mbp \ge \mbs$. The maximum masses of the primary and secondary
BHs are $\sim 50 \msun$ in the L model. This is because of
combinations of common envelope evolution and PPI effects as described
above. In contrast, the maximum masses of the primary and secondary
BHs are $\sim 100~\msun$ in the M model. The maximum masses are
roughly determined by the maximum BH mass obtained through single star
evolution, $\sim 95~\msun$ (Figure~\ref{fig:remnantMass}). The maximum
BH mass obtained through binary star evolution is larger than obtained
through single star evolution. This reason is explained as follows. A
binary system containing a $95~\msun$ star as a secondary star.  This
star evolves to a post-MS star without binary interaction. Its He core
mass is $< 45~\msun$. When it is a post-MS star, it gets extra mass
through stable mass transfer from its companion star, and grows to a
post-MS stars with $\sim 100~\msun$. Its He core mass does not
increase through stable mass transfer because it is a post-MS
star. Thus, it collapses to form a $\sim 100~\msun$ BH.

We focus on the results obtained using the L+K21 model. As seen in
Figures~\ref{fig:delayTime} and \ref{fig:massDist}, GW190521-like
BH-BHs are formed in the L+K21 model despite that stars partially
evolve along with the L model. Further, we describe the formation
mechanism of BH-BHs with the maximum BH mass of $\sim 80~\msun$,
similar to GW190521. A star with $\mzams \sim 90~\msun$ raises its He
core with $\lesssim 45~\msun$ when it enters into the post-MS
phase. Subsequently, it expands to $\gtrsim 10^3 \rsun$, and starts
stable mass transfer at a certain time. The stable mass transfer
increases $\mc/\mt$ of the post-MS star. When its $\mc/\mt$ becomes
larger than $\qcrit (= 0.58)$ (i.e. $\mc/\mt > \qcrit$), it suddenly
shrinks to its He core size, and stops the stable mass transfer. It
does not lose its mass until it collapses to a BH. Its He core mass is
$\lesssim 45~\msun$; thus, it can avoid PPI and PISN, and can directly
collapse to form a BH with $\sim 45 \qcrit^{-1}~\msun$, which is equal
to $78~\msun$. The maximum BH mass is also consistent with the result
of K21.

\begin{figure}
  \includegraphics[width=\columnwidth]{\fdir/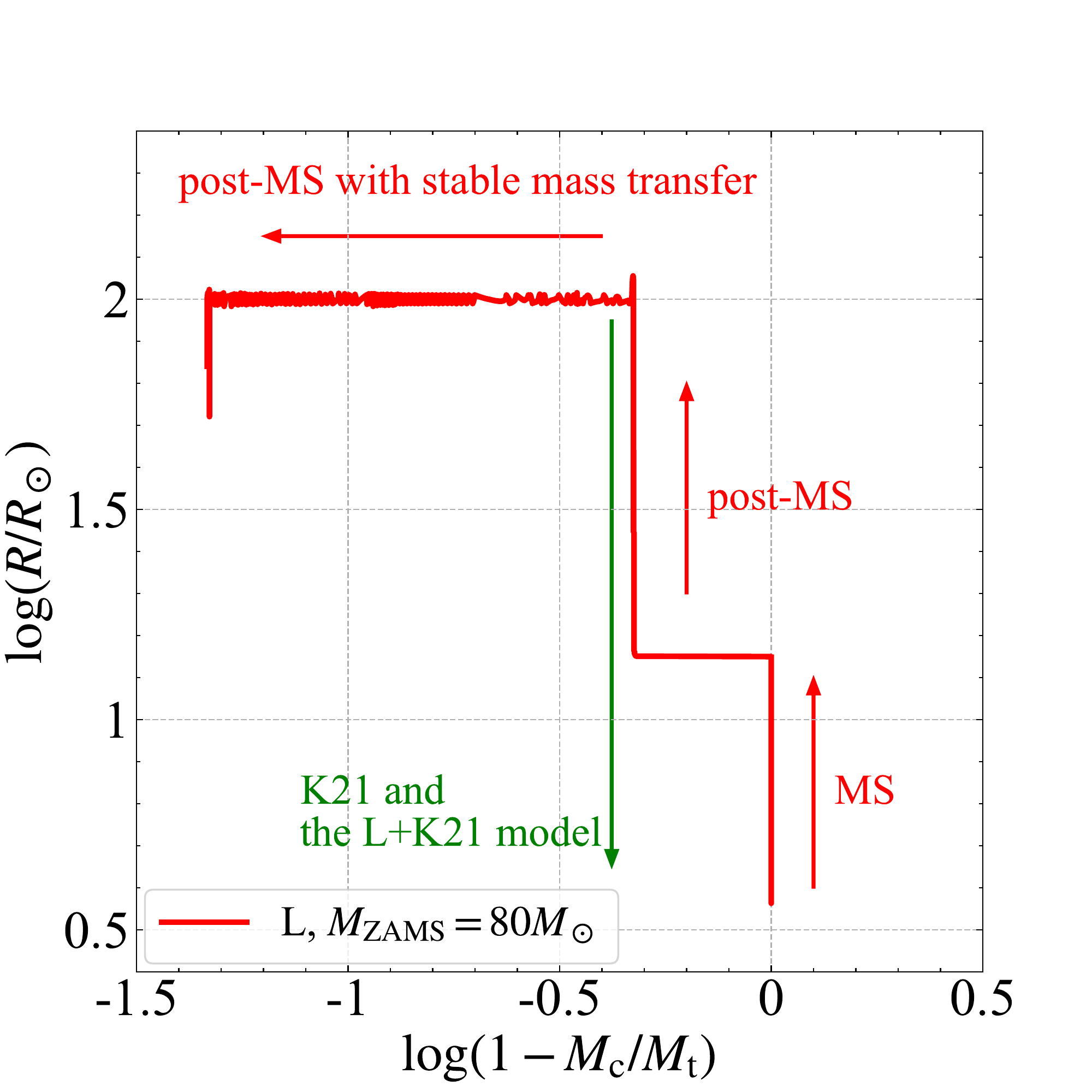}
  \caption{Radius evolution of a star with $\mzams =80~\msun$ in the L
    model with the He core mass fraction ($\mc/\mt$). We model the
    stable mass transfer such that the star loses its mass when it
    exceeds $10^2~\rsun$. The star has no He core in the MS phase;
    thus, $\mc/\mt = 0$ initially. In the post-MS phase, the star
    first expands freely up to $10^2~\rsun$. Subsequently, the star
    loses its mass through stable mass transfer, and maintains its
    radius $10^2~\rsun$ until $\mc/\mt \sim 0.95$. This means that
    $\qcrit \sim 0.95$ is correct for this star. If $\qcrit = 0.58$
    were correct, the star would shrink to its He core size at
    $\mc/\mt = 0.58$, as indicated by the green arrow.}
  \label{fig:radiusEvol}
\end{figure}

The comparison between the L and L+K21 models clearly shows that the
K21's modeling of post-MS stellar radii plays an important role in
forming GW190521-like BH-BHs. We examined whether $\qcrit = 0.58$ can
be applicable to Pop~III stars with $\mzams = 60 - 90~\msun$ in the L
model. We prepared a Pop~III star with $\mzams = 80~\msun$, which can
leave a BH in the PI~MG if it maintains its hydrogen envelope. The
star evolves in the L model by the {\tt HOSHI} code, a 1D stellar
evolution code \citep{2016MNRAS.456.1320T, 2018ApJ...857..111T,
  Takahashi19, Yoshida19}. The star experiences mass loss through
stable mass transfer so as not to exceed
$10^2~\rsun$. Figure~\ref{fig:radiusEvol} shows its radius
evolution. When the star enters into the post-MS phase, its He core
mass ($\mc$) is about half the total mass ($\mt$). The star exceeds
$10^2~\rsun$ at a certain time in the post-MS phase.  Subsequently, it
loses its hydrogen envelope, and $\mc/\mt$ increases. We can see that
the star keeps its radius $10^2~\rsun$ until $\mc/\mt \sim 0.95$. When
its $\mc/\mt$ exceeds $0.95$, the star suddenly shrinks. This results
mean $\qcrit \sim 0.95$ and are in favor of $\qcrit \sim 0.99$,
i.e. the L model. In contrast, K21 have modeled the response of a
stellar radius, such that a star shrinks down to its He core radius
($\sim 1~\rsun$) when $\mc/\mt = 0.58$ as indicated by the green arrow
in Figure~\ref{fig:radiusEvol}. This result shows that $\qcrit = 0.58$
is not applicable to a Pop~III star with $\mzams = 80~\msun$ in the L
model.

In summary, Pop~III binary systems in the M model (a model with a
small convective overshoot parameter) can form BH-BHs in the lower
half of the PI~MG ($65-100~\msun$). Consequently, BH-BHs in the M
model also cover well the error range of the masses of GW190521. In
contrast, BH-BHs in the L model (a model with a large convective
overshoot parameter) cannot form the mass combination of GW190521. In
the L+K21 model, Pop~III binary systems can form GW190521-like BH-BHs
because of their $\qcrit$. However, their $\qcrit$ is not applicable
to a Pop~III stars with $\mzams = 80~\msun$ which form BHs in the
PI~MG.

\section{Summary and Discussion}
\label{sec:Summary}

We perform BPS calculations, adopting the M and L models for Pop~III
star evolution models, where the M and L models have small and large
convective overshoot parameters. The M model is similar to the F21's
model, and the L model is the same as the T21's model. Additionally,
we treat the L+K21 model similar to the K21's model.

In the M model, Pop~III binary systems can form GW190521-like
BH-BHs. Stars in the M model maintain small radii until they
experience supernovae or direct collapse to BHs, and tend not to
interact with their companion stars. Finally, they can leave BHs in
the lower half of the PI~MG ($65-100~\msun$) through binary
evolution. The merger rate of BH-BHs in the lower half of the PI~MG is
$4 \times 10^{-2}$~$\yrgpc$ even in a pessimistic Pop~III formation
model. This is comparable to the merger rate of GW190521-like
BH-BHs. Consequently, Pop~III binary systems can be the origin of
GW190521 if the M model can apply to Pop~III star evolution. This
result is consistent with the F21's results.

In the L model, Pop~III binary systems cannot form GW190521-like
BH-BHs. Stars expand up to $\gtrsim 10^3~\rsun$, and actively interact
with their companion stars. They lose their hydrogen envelopes through
stable mass transfer or common envelope evolution, and become naked He
stars. Finally, they leave BHs with $\sim 45~\msun$ at most. Thus, the
choice of a convective overshoot parameter strongly affects the
formation of GW190521-like BH-BHs. We cannot conclude that the origin
of GW190521 is a Pop~III binary because determining the correct
overshooting parameter is highly uncertain.

In the L+K21 model, GW190521-like BH-BHs are also formed because of
their $\qcrit$. However, their $\qcrit$ is not reasonable for a
Pop~III star with $\mzams = 80~\msun$ that form BHs in the PI~MG.

Furthermore, we cannot also claim that the presence of GW190521 proves
the correctness of a model with a small convective overshoot parameter
(the M model) because there are many formation scenarios of GW190521,
which are described in section~\ref{sec:Introduction}. We expect
future GW observations to determine if a Pop~III binary system can be
the origin of GW190521. Clearly, Pop~III binary systems cannot form
the upper half of the PI~MG ($100-135~\msun$) through pure binary
evolution even in a model with a small convective overshoot parameter
(the M model), as shown in Figure~\ref{fig:massDist}. If the BH-BH
merger rate is suddenly decreased from the lower half of the PI~MG to
the upper half of the PI~MG, a Pop~III binary can be the origin of
GW190521\footnote{\cite{2021MNRAS.504..146V} have claimed that Pop~II
  stars can also form BHs with $\lesssim 100~\msun$ if the convective
  overshoot is ineffective. If this is the case, we may not identify
  if the origin of GW190521 is Pop~II or Pop~III stars, even if the
  BH-BH merger rate is suddenly decreased from the lower half of the
  PI~MG to the upper half of the PI~MG.}. Other formation scenarios
should have BH mass distributions with gradual decrease in the
ascending order of BH masses rather than BH mass distributions with
sudden decrease. These scenarios form BHs in the PI~MG through BH/star
mergers or gas accretion; thus, there is no reason for such sudden
decrease.

Figure~\ref{fig:massDist} shows a peak of secondary BHs at $\sim
45~\msun$. This is because PPI sweeps BH progenitors with He core mass
of $45-65~\msun$ to BHs with $45~\msun$. It appears that a Pop~III
binary tends to leave a BH-BH with a BH in the lower half of the PI~MG
and a BH with $\sim 45~\msun$. However, stars experiencing PPI do not
always leave BHs with $\sim 45~\msun$, and can leave BHs with $\sim
40-60~\msun$ in reality \citep{2019ApJ...887...72L,
  2020A&A...636A.104B}. Thus, we expect that a peak of secondary BHs
around at $\sim 45~\msun$ should be more mild, and cannot be a clue to
identify the origin of GW190521-like events.

\section*{Acknowledgments}

We are grateful to Alessandro A. Trani for checking our manuscript,
and anonymous referee for many suggestions for improving this
paper. This research was supported in part by Grants-in-Aid for
Scientific Research (17H01130, 17H06360, 17K05380, 19K03907, and
20H05249) from the Japan Society for the Promotion of Science, and by
University of Tokyo Young Excellent researcher program. We would like
to thank Editage (www.editage.com) for English language editing.

\section*{Data availability}

Results will be shared on reasonable request to authors.

\appendix

\section{The critical mass fraction in {\tt BSE}}
\label{sec:qcrit}

In this section, we show $\qcrit \sim 0.99$ in {\tt SSE/BSE}
\citep[][respectively]{2000MNRAS.315..543H,
  2002MNRAS.329..897H}. \cite{2000MNRAS.315..543H} describe the radius
of a post-MS star with large $\mc/\mt$ in section 6.3 of their
study. A post-MS star with large $\mc/\mt$ has $\mu < 1$, where $\mu$
is defined as follow:
\begin{align}
  \mu = \left( 1- \frac{\mc}{\mt} \right) \min \left\{ 5.0, \max\left[
    1.2, \left( \frac{L}{7 \times 10^4 \lsun} \right)^{-0.5} \right]
  \right\}. \label{eq:ratio}
\end{align}
The radius of this type of post-MS star is expressed as follows:
\begin{align}
  R &= \rcore \left( \frac{\rpost}{\rcore}
  \right)^\rho \label{eq:radius} \\
  \rho &= (1+c^3) \frac{(\mu/c)^3}{1+(\mu/c)^3}
  \mu^{0.1/\ln(\rpost/\rcore)} \label{eq:power} \\
  c &= 0.006 \max \left( 1, \frac{2.5~\msun}{\mt}
  \right), \label{eq:constant}
\end{align}
where $\rpost$ is the radius of a post-MS star with $\mu \ge 1$ and
$\rcore$ is the radius of a post-MS star without hydrogen envelope.
Eqs.~ (\ref{eq:ratio}), (\ref{eq:radius}), (\ref{eq:power}), and
(\ref{eq:constant}) correspond to Eqs.~(97), (100), (102), and (104),
respectively. Note that these notations are different from those used
by \cite{2000MNRAS.315..543H}.

Eq.~(\ref{eq:radius}) presents $R \rightarrow \rcore$ for $\rho
\rightarrow 0$. Let us consider how $\rho$ decreases with decrease in
$\mu$. $\mt > 10~\msun$ for our case; thus, $c=0.006$, and
$(1+c^3) \sim 1$. Furthermore, $\rpost/\rcore \sim 10^2-10^4$,
$0.1/\ln(\rpost/\rcore) \lesssim 0.02$, and thus,
$\mu^{0.1/\ln(\rpost/\rcore)} \sim 1$ for $\mu \sim c$. In contrast,
$(\mu/c)^3/[1+(\mu/c)^3] \gtrsim 0.999$ for $\mu \gtrsim 10c$, and
$(\mu/c)^3/[1+(\mu/c)^3] \sim 0.5$ for $\mu \sim c$. Consequently,
$\rho$ suddenly decreases for $\mu \sim c$. The $\min$ term in
Eq.~(\ref{eq:ratio}) is equal to $1.2$ for our case; thus, $\mc/\mt
\sim 0.95$ and $0.995$ for $\mu \sim 10c$ and $c$,
respectively. Therefore, we can see $\qcrit \sim 0.99$ in {\tt BSE}.

\section{Numerical noise of BPS calculations}
\label{sec:NumericalNoise}

\begin{figure}
  \includegraphics[width=\columnwidth]{\fdir/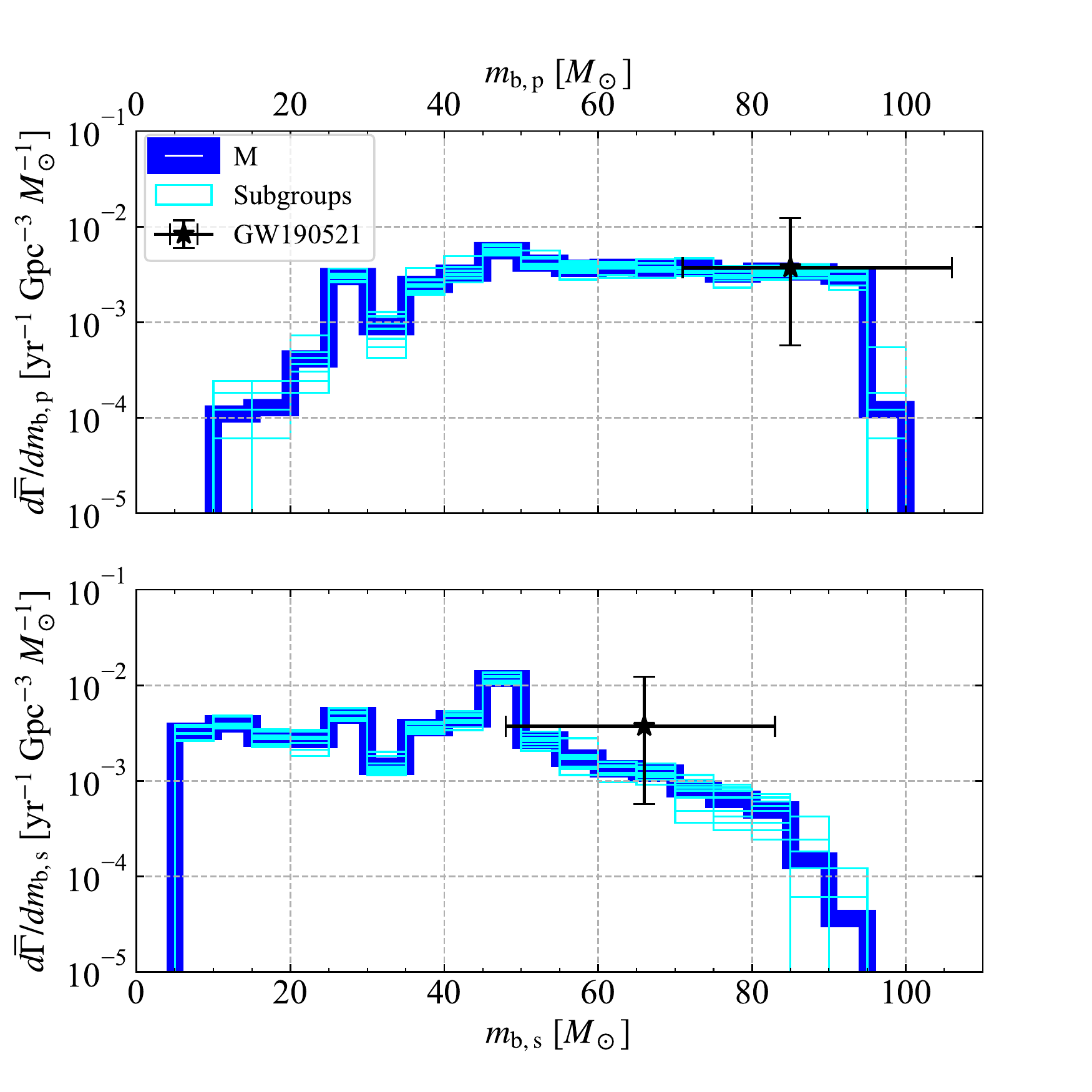}
  \caption{Mass distribution of the merging BH-BHs at the redshift of
    the GW190521 event in the M model and its $10$ subgroups. The
    upper and lower panels correspond to the upper and right panels of
    Figure~\ref{fig:massDist}, respectively. The thick blue curves and
    star marks with error bars are the same as those shown in
    Figure~\ref{fig:massDist}.}
  \label{fig:staticsMassDist}
\end{figure}

We estimated the numerical noise of BPS calculations by the jackknife
resampling. We obtained $10$ subgroups, dividing $10^6$ binary systems
into $10$ equal parts. Subsequently, we investigated the difference in
BH-BH properties among the $10$ subgroups of $10^6$ binary systems in
the M model. Figure~\ref{fig:staticsMassDist} shows the mass
distributions of BH-BHs in the M model and its subgroups. We can see
that the difference of mass distributions among these subgroups is the
thickness of the blue curves for $25 \lesssim \mbp/\msun \lesssim 95$
and $\mbs/\msun \lesssim 85$. For other mass ranges, the numerical
noise is a bit large because the number of merging BH-BHs is small in
BPS calculations. Nevertheless, the numerical noise is negligible when
we compare our results with GW190521.

%\bibliographystyle{mnras}
%\bibliography{natbib} % if your bibtex file is called example.bib

\bsp	% typesetting comment
\label{lastpage}
\end{document}